\documentclass[12pt]{article}
\def\href{\relax}
\usepackage{subfigure} \usepackage{cite}
\usepackage{citesort}
\usepackage{latexsym} \usepackage{amssymb} \textwidth 16cm
\usepackage{amsmath} \usepackage{pictex} \usepackage{graphics}
\newcommand{\eq}[1]{Eq.~(\ref{#1})}

\begin{document}
\title{Polymer dynamics  in time-dependent Matheron-de Marsily flows:
An exactly solvable model.}  \author{S. Jespersen,$^{a,b}$
G. Oshanin,$^{b,c}$ and A. Blumen$^{b}$\\ $^a$Institute of Physics and
Astronomy\\ University of Aarhus, DK-8000 {\AA}rhus C, Denmark\\
$^b$Theoretische Polymerphysik\\ Universit\"at Freiburg, D-79104
Freiburg i.Br., Germany\\ $^c$Laboratoire de Physique des Liquides\\
Universit\'e Paris IV, 4, Place Jussieu, F-75252 Paris, France.}

\date{\today}   \maketitle
\begin{abstract}
We introduce a new model of random layered media, extending the
Matheron-de Marsily model: Here we allow for the flows to change in
time. For such layered structures, we solve exactly the equations of 
motion for single particles, and also for polymers modelled as Rouse
chains. The results show a rich variety of dynamical patterns.
\end{abstract}
\parbox{1\linewidth}{PACS numbers: 36.20.-r, 47.55.-t, 05.60.-k}
\section{Introduction}

The dynamics of particles and particle assemblies in random force
fields  is a subject of intense current interest (see
Ref.\cite{gleb1,gleb2,wiese} 
and references therein).  Such fields often lead to qualitative
deviations from simple dynamic patterns, and result in the anomalous
diffusion of the particles involved.  A standard prototype model is
that of Matheron and de Marsily flows (MdM)
\cite{gleb1,gleb2,wiese,mdm,redner,zumofen1,zumofen2,bouchaud1,bouchaud2,doussal},
originally designed to 
describe the  transport of a solute in porous media. Similar
mathematical forms arise when treating electrons in random potentials
or  spin depolarization in random fields\cite{doussal}.  In the MdM model
the environment is viewed as  consisting of  layers of force fields
\cite{mdm}.  The particles' mean square displacement (msd) in the
direction parallel to the flows, say, along the $Y$-axis,  $<Y^2(t)>$,
can be evaluated exactly;  it has been shown that it grows  as
$<Y^2(t)> \sim t^{3/2}$ \cite{mdm,bouchaud1,bouchaud2}, $t$
being the time.  Recently the dynamics of  Rouse polymer chains in
time-independent MdM flows were also discussed and several exact
results have been derived \cite{gleb1,gleb2}. In particular,  it has
been shown that the msd of some tagged bead of the chain may display
different dynamical regimes depending on whether the time of
observation is shorter or longer than the so-called Rouse time
$t_R$. Now $t_R$ goes as $N^2$, where $N$ is the degree of
polymerisation, i.e. the chain's length  \cite{rouse,doi,degennes}.  For
times shorter than $t_R$ a sub-ballistic law $<Y^2(t)> \sim t^{7/4}$
was derived, while it was shown that for times longer than $t_R$ the
$t^{3/2}$-dependence is restored, the prefactor being a growing
function of $N$.  Thus a tagged monomer of a polymer immersed in MdM flows
moves in general faster than an individual particle (i.e. a chain with $N = 1$),
a fact which has its physical explanation\cite{gleb1,gleb2}.

These results concern, however, velocity fields whose directions and
magnitudes are random in space but fixed in time: we speak of quenched
disorder. For random divergenceless velocity flows, for which
the flows change with time, we show in the following that new results
emerge. 

In the present paper we present exact dynamical solutions both for 
individual particles and for long Rouse chains moving in
time-dependent, random layered media.  Distinct from the previous works
\cite{mdm,bouchaud1,bouchaud2,gleb1,gleb2}, we now allow the
velocities in the layers to change randomly with time. We thus take
into account the fact that the environment itself may be subject to
dynamical processes, thereby introducing new time-scales into the
problem. By this the time fluctuations of the flow give rise to
interesting, novel dynamical 
behaviors. In particular, we show that at intermediate times ($t \ll
t_R$) the msd of any tagged bead of a polymer chain follows  $<Y^2(t)> \sim
t^{3/4}$, i.e. the msd moves faster than the Rouse law
$<Y^2(t)> \sim t^{1/2}$ 
which holds in the flow-free case \cite{doi,degennes}. On the other
hand, at times $t \gg t_R$, we recover a diffusive behavior, 
in which however, the effective diffusion constant scales
as $N^{-1/2}$ for sufficiently long chains, 
i.e. vanishes with $N$ at a slower rate than the Rouse diffusion
constant, which goes as $1/N$.

The paper is structured as follows: In Section II we formulate the model
and introduce the basic notation. In Section III we analyse the dynamics of
an individual particle subject to time-dependent MdM
flows. Section IV is devoted to the analysis of the dynamics of a
tagged bead of an infinitely long Rouse chain, while Section V deals
with effects typical for finite chains. We conclude with a discussion
and a brief summary of results in Section VI.   

\section{The model}\label{intro}

A standard polymer model due to Rouse \cite{rouse} consists in
viewing the macromolecule as a series of $N$ beads, linearly connected
by harmonic 
springs. Its dynamics in solution, but subject to external force
fields and excluding hydrodynamic interactions and steric hindrances,
is given by the  Langevin equation  \cite{doi,degennes,rouse}:
\begin{equation}
\label{langevin}
\zeta\frac{d {\mathbf R}_n}{d t}=K \Big({\mathbf R}_{n+1} +  {\mathbf
R}_{n-1} - 2 {\mathbf R}_n \Big) +{\mathbf F}({\mathbf
R}_n,t)+\boldsymbol{\eta}(n,t),
\end{equation}
for $1 \leq n \leq N$, complemented by ${\mathbf R}_{0} \equiv
{\mathbf R}_{1}$ and ${\mathbf R}_{N} \equiv {\mathbf R}_{N+1}$.  In
this equation 
$\zeta$ is the coefficient of friction, ${\mathbf R}_n$ is the
position of the $n$th monomer, $K=3k_B T/b^2$ is the spring constant accounting
for the Hookean interaction of the monomers, $k_B T$ being the
temperature multiplied by the Boltzmann constant and $b$ the average
distance between the beads, and $\boldsymbol{\eta}(n,t)$ represents
the thermal noise due to 
interactions with the solvent.  In Eq.(\ref{langevin}) the ${\mathbf
F}({\mathbf R}_n,t)$ denote the extra forces due to the imposed
velocity fields. Equation~(\ref{langevin}) is a simplified
description of the dynamics of a polymer in  $\theta$-solutions
\cite{degennes,doi}. 

Treating the index $n$ as continuous leads to the replacement of the
discrete term $\Big({\mathbf R}_{n+1} +  {\mathbf R}_{n-1} - 2
{\mathbf R}_n \Big)$  by  the Laplacian operator  $\partial^2 {\mathbf
R}_n/ \partial n^2$ and of  ${\mathbf R}_{0} \equiv {\mathbf R}_{1}$
and ${\mathbf R}_{N} \equiv {\mathbf R}_{N+1}$ by the  Rouse boundary
conditions\cite{doi}:
\begin{equation}
\label{bc}
\frac{\partial {\mathbf R_n}}{\partial n}=0, \;\;\; \text{for $n=0$
and $n=N$}.
\end{equation}

Next, we specify the forces entering the right-hand-side (rhs) of
Eq.(\ref{langevin}).  We let, as usual, the thermal noise be
Gaussian with zero mean; due to the fluctuation-dissipation theorem
the second moment takes the form:
\begin{equation}
\label{thermal}
\overline{\eta_\alpha(n,t)\eta_\beta(n',t,)}=2k_B
T\zeta\delta_{\alpha,\beta}\delta_{n,n'}\delta(t-t'),
\end{equation}
In Eq.(\ref{thermal}) we let the Greek indices denote the cartesian
components.  The Matheron-de Marsily (MdM) model is obtained by
assuming layered flows, say by taking the
forces to be along the $Y$-axis, but to depend solely on the
$X$-component, see e.g. Fig.~(1) in \cite{gleb1}:
\begin{equation}
{\mathbf F}({\mathbf R}_n,t)=(0,f(X_n,t),0).
\end{equation}
This choice of ${\mathbf F}$ models a layered medium along the
$X$-axis; each layer has a random (but for all points of the layer
fixed)  velocity pointing along the $Y$-axis.  Due to the uncoupling
of  the different components in the Rouse-model, the dynamics in the
transversal plane, (i.e.  the $X$- and $Z$-components) are not
influenced by the presence of  the velocity fields. However, as we
shall see, the motion parallel to the $Y$-axis is dramatically changed.

 The random force ${\mathbf F}({\mathbf R}_n,t)$, (i.e. its
$Y$-component)  is assumed to be Gaussian and zero-centered. For the
second moment we take
\begin{equation}
\label{environment}
\langle f(X,t)f(X',t')\rangle=\frac{\Delta}{2\Gamma} \delta(X-X')
e^{-|t-t'|/\Gamma}
\end{equation}
Note that the strength of the random velocity field depends on the
constants  $\Gamma$ and $\Delta$.  Equation~(\ref{environment})
defines a changing environment with a short term memory, exponentially
decreasing  on the time scale $\Gamma$;  we will call $\Gamma$ the
renewal time. In the limit $\Gamma\rightarrow  \infty$ and keeping the
ratio $\Delta/\Gamma$ fixed we recover the standard quenched MdM
model. We shall denote this limiting procedure as the {\it quenched
limit} in the rest of the paper.  On the other hand, when
$\Gamma\rightarrow 0$, $\exp(-|t|/\Gamma)/(2\Gamma)$ tends to the
Dirac $\delta(t)$ distribution; this limit allows us to study  an
environment without memory.  We distinguish averages over thermal
histories from averages over configurations of the velocity field by
using an overbar for the former and angular brackets for the latter.

\section{Dynamics of a single particle}

To fix the ideas, we start with the case of a  single  bead in
time-dependent MdM flows. Then $ N = 1$ and $K = 0$. Because  of the
decoupling of the different coordinates, the $X$- and $Z$-components of
the particle's displacement obey the force-free Langevin equation
\begin{equation}
\label{langxz}
\zeta\frac{dX}{dt}=\eta_x(t),
\end{equation}
with  $\eta_x(t)$ being the $X$-component of the thermal noise.
Equation (\ref{langxz}) yields for the  mean square displacement (msd)
averaged over thermal histories the usual diffusion result: 
\begin{equation}
\label{diffusion}
\overline{X^2(t)} =2\frac{k_B T}{\zeta}=2Dt,
\end{equation}
Through the last relation we have introduced the ``bare'' diffusion
constant $D$.  For the $Y$-component we have to solve the equation
\begin{equation}
\label{langy}
\zeta\frac{dY}{dt}=\eta_y(t)+f(X(t),t)
\end{equation}
with $f(X(t),t)$ being the random function of Eq.(\ref{environment}).
The solution of Eq.(\ref{langy}) is:
\begin{equation}
Y(t)=\frac{1}{\zeta}\int_0^t d\tau(\eta_y(\tau)+f(X(\tau),\tau)),
\end{equation}
where we set $Y(0)=0$. Being interested in the mean square
displacement we have:
\begin{equation}
\langle \overline{Y^2(t)}\rangle=2Dt+\frac{1}{\zeta^2}\int_0^t
dt_1\int_0^t dt_2  \langle
\overline{f(X(t_1),t_1)f(X(t_2),t_2)}\rangle,
\end{equation}
by noting that the different sources of randomness are decoupled.  Then,
in virtue of \eq{environment} we have for the second term 
\begin{equation}
\label{int}
\int_0^t dt_1\int_0^t dt_2  \langle
\overline{f(X(t_1),t_1)f(X(t_2),t_2)}\rangle=\frac{\Delta}{2\Gamma}\int_0^t
dt_1\int_0^t dt_2 \overline{\delta(X(t_1)-X(t_2))}
e^{-|t_1-t_2|/\Gamma}.
\end{equation}
Now it is a simple matter to average the delta-function on the
rhs of Eq.(\ref{int}), by making use of its Fourier
representation; this yields:
\begin{eqnarray}
\label{delta}
\nonumber
\overline{\delta(X(t_1)-X(t_2))}&=&\int_{-\infty}^{\infty}\frac{dk}{2\pi}
\overline{\exp\left(-ik(X(t_1)-X(t_2))\right)}\\ \nonumber
&=&\int_{-\infty}^{\infty}\frac{dk}{2\pi}\exp\left(-Dk^2|t_1-t_2|\right)\\
&=&\frac{1}{\sqrt{4\pi D|t_1-t_2|}}.
\end{eqnarray}
Here we noted that the process $X(t)$ is Gaussian and  we used in the
second line the well-known rule for averaging Gaussian
exponential forms.  Inserting Eq.(\ref{delta}) into \eq{int} and
integrating leads to:
\begin{equation}
\label{msd}
\langle\overline{Y^2(t)}\rangle=2 D t +\frac{\Delta}{2\zeta^2}
\sqrt{\frac{\Gamma}{D}}\left((t/\Gamma-1/2)\,{\mbox{erf}}\,[\sqrt{t/\Gamma}]+\sqrt{t/(\pi\Gamma)}e^{-t/\Gamma}\right),
\end{equation}
where ${\mbox{erf}}\,[x]$ denotes the error-function, Eq.(7.1.1) in
Ref.\cite{abr}.  Equation~(\ref{msd}) is exact and is the main result
of this section.  From \eq{msd} we have in the limit $t \ll \Gamma$:
\begin{equation}
\label{tsmall}
\langle\overline{Y^2(t)}\rangle=2Dt+\frac{2\Delta \Gamma^{1/2}}{3
\zeta^2\sqrt{D\pi}}\, \Big( \frac{t}{\Gamma}\Big)^{3/2}+ {\cal O}\Big(\Big(
\frac{t}{\Gamma}\Big)^{5/2}\Big), 
\end{equation}
where the second term represents the result for standard, quenched
MdM flows \cite{mdm}.  The physics underlying such a behavior is
well understood\cite{gleb1,gleb2,mdm,redner,zumofen1,zumofen2} and it
is due to dynamically induced correlations in the velocities felt by
the particle.  

In the opposite limit,  $t \gg \Gamma$, we are lead to
\begin{equation}
\label{eq}
\langle\overline{Y^2(t)}\rangle = 2 D t +\frac{\Delta}{2
\zeta^2\Gamma}\sqrt{\frac{\Gamma}{D}} t + {\cal{O}}(1) = 2D_{eff} t +
{\cal{O}}(1),  
\end{equation}
where
\begin{equation}
\label{Deff}
D_{eff}=D+\frac{\Delta}{4\zeta^2\Gamma}\sqrt{\frac{\Gamma}{D}},
\end{equation}
i.e. we find diffusion with a renormalized diffusion coefficient.

Thus for large times $t$ the fluctuations of the environment (flow)
contribute additively to the diffusion coefficient. Note, however,
that now   diffusion is anisotropic: Parallel  to and perpendicular to
the layers one has $D_{||}=D_{eff}$ and $D_{\perp}=D$,
respectively. The  dependence of $D_{eff}$ on both $D$ and $\Gamma$ is easily
visualized: For larger $D$ the perpendicular motion is more
rapid and hence the velocity field viewed by the particle changes more
rapidly (and is less efficient); furthermore, a decrease in $\Gamma$
leads to a similar effect. On the other hand, in the $\Gamma \to \infty$ limit
$D_{eff}$ diverges, a sign that the motion gets to be superdiffusive,
see \eq{tsmall}.  

Since the disorder changes completely on the time scale $\Gamma$, one
may view Eq.(\ref{eq}) as arising from renewals of the process in
Eq.(\ref{tsmall})  (with total loss of memory) every
$\tau$ units of time, where $\tau$ is of the order of
$\Gamma$. In fact, since the crossover in behavior 
from $(2 \Delta \Gamma^{1/2}/3 \zeta^{2} 
(\pi D)^{1/2}) (t/\Gamma)^{3/2}$ to $(\Delta/2 \zeta^{2} (\Gamma
D)^{1/2}) t$ occurs at $t_c = 9 \pi \Gamma/16 \approx 1.767 \Gamma$,
we reproduce \eq{Deff} exactly if we take $\tau=t_c$.

\section{Infinitely Long Rouse Chain}

We consider next the opposite case of an infinitely long Rouse chain.
For a given thermal history, the solution of \eq{langevin} for the
$Y$-component of the displacement of the  $n$-th bead can be readily
found by standard means, see Ref.\cite{gleb1,gleb2}, and reads
\begin{equation}
\label{solx}
Y_n(t)=\frac{1}{\zeta}\int_0^t d\tau\,\int_{-\infty}^\infty dl\,
P(n-l,t-\tau)(f(X_l(\tau),\tau)+\eta_y(l,\tau)),
\end{equation}
where we again, for simplicity, assumed that  $Y_n(0)=0$. Here
$P(l,\tau)$ is the Green's function solution of the $1D$ diffusion
equation 
\begin{equation}
\label{kernel}
P(l,\tau)=\sqrt{\frac{\zeta}{4\pi K\tau}}\exp\left[-\frac{\zeta
    l^2}{4K\tau}\right].
\end{equation}
The msd of the $n$-th bead, averaged both over thermal histories and
over random velocity fields, takes the form:
\begin{multline}
\label{r}
\langle\overline{Y^2_n(t)}\rangle=2D\left(\frac{\zeta}{2\pi
    K}\right)^{1/2} t^{1/2}+ \frac{\Delta}{2\Gamma\zeta^2}\int_0^t
    d\tau_1 \int_0^t d\tau_2 \int_{-\infty}^{\infty}
    dl_1\int_{-\infty}^{\infty} dl_2 P(n-l_1,t-\tau_1) \times \\
    \times P(n-l_2,t-\tau_2)\, e^{-|\tau_1-\tau_2|/\Gamma}
    \,\overline{\delta(X_{l_1}(\tau_1)-X_{l_2}(\tau_2))}
\end{multline}
where we used the second moment of the velocity field, given by
\eq{environment}. 

To proceed further, we need to know
$\overline{\delta(X_{l_1}(\tau_1)-X_{l_2}(\tau_2))}$.  This form  can
be evaluated exactly (see  Refs.\cite{gleb1,gleb2,degennes,doi}
for more details). The result  (for $\tau_1 \leq \tau_2$) is
explicitly: 
\begin{equation}
\label{functional}
\overline{\delta(X_{l_1}(\tau_1)-X_{l_2}(\tau_2))}=\left(\frac{K}
{4\pi\zeta D^2}\right)^{1/4} M(l_2-l_1,\tau_1,\tau_2)^{-1/2},
\end{equation}
in which the function $M(l,\tau_1,\tau_2)$ stands for
\begin{equation}
\label{M}
M(l,\tau_1,\tau_2)=(2\tau_1)^{1/2}+(2\tau_2)^{1/2}-4\left(\frac{\pi
    K}{\zeta}\right)^{1/2} \int_0^{\tau_1}d\tau
    P(l,\tau_1+\tau_2-2\tau)\
\end{equation}
Inserting Eqs.(\ref{functional}) and (\ref{M}) into Eq.(\ref{r}) and
reverting to dimensionless variables, say, by setting  
\begin{gather}
\nonumber \theta_1=\tau_1/t; \hspace{.5cm}\theta_2= \tau_2/t;
\hspace{.5cm}\theta=\tau/t;  \\  z_1=
(l_1-n)\displaystyle{\left(\frac{\zeta}{4Kt}\right)^{1/2}};
\hspace{.5cm} z_2=
(l_2-n)\displaystyle{\left(\frac{\zeta}{4Kt}\right)^{1/2}} 
\end{gather}
we finally arrive at the following expression:
\begin{equation}
\label{bead}
\langle\overline{Y^2_n(t)}\rangle=2D\left(\frac{\zeta}{2\pi
    K}\right)^{1/2} t^{1/2}
    +\frac{\Delta}{\pi\Gamma\zeta^2}\left(\frac{K}{4D^2\pi\zeta}\right)^{1/4}
    t^{7/4} g(t/\Gamma),
\end{equation}
with $g(\eta)$ being the dimensionless function:
\begin{equation}
\label{g}
g(\eta)=\int_0^1d\theta_2\int_0^{\theta_2}d\theta_1 \;
e^{-(\theta_1-\theta_2) \eta} \int_{-\infty}^{\infty}dz_1
\int_{-\infty}^{\infty}dz_2\,\frac{e^{-z_1^2/(1-\theta_1)
-z_2^2/(1-\theta_2)}}
{\sqrt{(1-\theta_1)(1-\theta_2)\tilde{M}(z_1-z_2,\theta_1,\theta_2) }}.
\end{equation}
where for $\theta_2\geq\theta_1$ the function
$\tilde{M}(z,\theta_1,\theta_2)$ is 
\begin{equation}
\label{mtilde}
\tilde{M}(z,\theta_1,\theta_2)=\sqrt{2\theta_1}+\sqrt{2\theta_2}-2\int_0^{\theta_1}
d\theta\,
\frac{e^{-z^2/(\theta_1+\theta_2-2\theta)}}{\sqrt{\theta_1+\theta_2-2\theta}}.
\end{equation}
In the limit $\Gamma \to \infty$ (quenched MdM flows) $\eta =
t/\Gamma$ tends to zero and  $g(\eta)$ to a constant.  Hence, for
$\Gamma \to \infty$ and not-too-small $t$ one has
$\overline{<Y_n^2>}\sim t^{7/4}$; this reproduces precisely the
corresponding result  derived in Refs.\cite{gleb1} and \cite{gleb2}.
Moreover, we expect   this dynamical behavior  to show up also for
finite $\Gamma$, as long as $t \ll \Gamma$, i.e. $\eta \ll 1$.

Turning now to finite but small $\Gamma$, for $t \gg \Gamma$, i.e.
$\eta = t/\Gamma \gg 1$,  the leading large-$\eta$ behavior of
$g(\eta)$ can be found by integrating in Eq.(\ref{g}) over $\theta_1$
by parts with respect to $exp(\theta_1 \eta)$.  We find  that
$g(\eta)$ has the form:
\begin{equation}
\label{g2}
g(\eta)= \frac{C}{\eta} + {\cal{O}}(\frac{1}{\eta^2}), 
\end{equation}
where $C$ is  a dimensionless constant, given by
\begin{equation}
\label{g3}
C=\int_0^1d\theta\int_{-\infty}^{\infty}dz_1
\int_{-\infty}^{\infty}dz_2
\,\frac{e^{-(z_1^2+z_2^2)/(1-\theta)}}{(1-\theta)
\sqrt{\tilde{M}(z_1-z_2,\theta,\theta)}}
\end{equation}
Inserting now Eq.(\ref{g2}) into \eq{bead} we find
that the msd of any bead of an infinitely long Rouse chain obeys
\begin{equation}
\label{bead2}
\langle\overline{Y^2_n(t)}\rangle= 2D\left(\frac{\zeta}{2\pi
    K}\right)^{1/2}\,t^{1/2}+
    \frac{C\Delta}{\pi\zeta^2}\left(\frac{K}{4D^2\pi\zeta}\right)^{1/4}
    t^{3/4} +{\cal{O}}(t^{-1/4}).
\end{equation}
The significant terms  of Eq.(\ref{bead2}) can be understood as
follows: The first term is just the Rouse-result in the absence of
flows. The second term is a new feature here, which arises due to the
time-dynamics of the MdM flows, to be contrasted with the
$t^{7/4}$-term for quenched MdM flows. This new term appears, of
course, even when the time-correlation in \eq{environment} is of
$\delta$-form, i.e. for $\Gamma\rightarrow 0$.

Therefore, we have for monomers attached to long polymer chains that
at long times  $<\overline{Y_n^2(t)}> \sim t^{3/4}$, i.e. that the exponent
of the anomalous diffusion  is larger than in the force-free case.
One may contrast this behavior to that of a single particle, where at
long times time-dependent velocity fields only change the diffusion
constant, but leave the exponent of $t$ (unity) unchanged. Hence the
influence of the velocity fields on the monomer's motion differs from
that of the thermal noise. The difference can be traced to the form of
the disorder; the solvent's role is included through MdM
$\delta(X_{l_1}(t_1)-X_{l_2}(t_2))$-fields, which depend on the
$X$-components of the positions of the beads, while thermal noise
depends on $\delta_{l_1,l_2}$.  This means that while different
monomers always experience different forces from the heat bath,
distinct monomers with the same $X$-components (there are many such
monomers, since the polymer's projection on the $X$-axis corresponds
to a simple random-walk) experience identical velocity fields, a fact which
qualitatively enlarges $<\overline{Y_n^2(t)}>$.

\section{Finite Rouse Chains}

For finite Rouse chains there appears yet another timescale, namely
the so-called Rouse time $t_R=\zeta N^2/(\pi^2K)$
\cite{rouse,doi,degennes}, $t_R$ being  the largest internal relaxation
time of the structure.  To display its role we start by solving
\eq{langevin} for finite $N$. The solution is found most readily in
terms of normal coordinates\cite{doi}, introduced, say for $Y_n(t)$,
through: 
\begin{equation}
\label{normal}
Y_n(t)=\sum_{p=-\infty}^{\infty}\cos(p\pi n/N) Y(p,t),
\end{equation}
and similarly for $X_n(t)$ and $Z_n(t)$. Inserting these forms into the
equation of motion, \eq{langevin}, and solving yields straightforwardly
for the $X$-component \cite{gleb1}
\begin{equation}
\label{xsolfinite}
X(p,t)=\frac{1}{N\zeta}\int_0^t
d\tau\int_0^Ndn\,\eta_x(n,\tau)\cos(p\pi n/N)\,  e^{-p^2(t-\tau)/t_R}
\end{equation}
and similarly for the $Z$-component. For the $Y$-component one has
instead 
\begin{equation}
\label{ysolfinite}
Y(p,t)=\frac{1}{N\zeta}\int_0^t
d\tau\int_0^Ndn\,\left(\eta_x(n,\tau)+f(X_n(\tau),\tau)\right)\cos(p\pi
n/N)\, e^{-p^2(t-\tau)/t_R}.
\end{equation}
The behavior of the beads depends on the ratio $t/t_R$. For short
times $t\ll t_R$ the sums over $p$ for $X_n(t)$, $Y_n(t)$ and $Z_n(t)$,
e.g. \eq{normal}, can be converted into integrals; exemplarily, inserting
\eq{ysolfinite} into \eq{normal} and performing the integration leads
to \eq{solx}. Thus for $t\ll t_R$ each bead behaves as if it were part of an 
infinite chain: Hence the results from the previous section
apply. In the opposite regime, for $t\gg t_R$, only
the zeroth mode $p=0$ contributes significantly to 
\eq{ysolfinite}, and the motion of each bead follows closely that of
the center of mass (CM) of the chain. In the case of the $X$- and
$Z$-components, this is a diffusive  
behavior with the renormalized diffusion constant $D\rightarrow
D/N$ \cite{doi,rouse,degennes}. For the $Y$-component, however, we have for $t\gg t_R$ approximately:
\begin{eqnarray}
\label{yp2}
&&\langle\overline{Y_n^2(t)}\rangle\simeq
 \langle\overline{Y^2(0,t)}\rangle= 2\frac{D}{N}t+\nonumber\\
&+&\frac{\Delta}{2N^2\Gamma\zeta^2}\int_0^t
d\tau_1\int_0^td\tau_2\int_0^Ndn_1\int_0^Ndn_2\,
\overline{\delta(X_{n_1}(\tau_1)-X_{n_2}(\tau_2))}
e^{-|\tau_1-\tau_2|/\Gamma} .
\end{eqnarray}
We remark that the terms on the rhs of \eq{yp2} describe the dynamics
of the $Y$-component of the chain's CM for all $t$. The evaluation of
\eq{yp2} is 
rendered complex for finite $N$ due to the 
appearance of the two-time delta function in the integral. It is,
however, possible to proceed along the lines of Ref.~\cite{gleb2}, but
this is outside the scope of the present article. Since the
analysis simplifies for $\Gamma\rightarrow \infty$ and for
$\Gamma\rightarrow 0$, we shall consider only these two cases in the
following.  

For $\Gamma\rightarrow \infty$ the flows get to be time-independent; then
the approach follows the derivation given in \cite{gleb1,gleb2} for
quenched MdM flows. There it was found that the leading behavior of
the bead's motion obeys (Eq.~(36a) of Ref.~\cite{gleb1}, here in our
notation): 
\begin{equation}
\label{quench}
\langle \overline{ Y_n^2(t)}\rangle\sim \frac{\Delta}{\Gamma\zeta^2}
\sqrt{\frac{N}{D}} t^{3/2},
\end{equation}
that is, the CM moves practically as an individual particle,
i.e. \eq{tsmall}, with $D$ being however replaced by $D/N$.  

For $\Gamma\rightarrow 0$, we have for the second term (call it $I$) in
\eq{yp2} 
\begin{equation}
\label{1timedelta}
I=\frac{\Delta}{N^2\zeta^2}\int_0^t d\tau\int_0^Ndn_1\int_0^Ndn_2\,
\overline{\delta(X_{n_1}(\tau)-X_{n_2}(\tau))}.
\end{equation}
For  $t\gg t_R$ most of the $\tau$ in \eq{1timedelta} also obey
$\tau\gg t_R$, so that we may view the Gaussian $X$-process to be 
stationary,
i.e. $\overline{X_{n_1}(\tau)-X_{n_2}(\tau)}=D\zeta|n_1-n_2|/K$,
independent of $\tau$. Hence
\begin{eqnarray}
\label{gauss}
\nonumber
\overline{\delta(X_{n_1}(\tau)-X_{n_2}(\tau))}&=&\int_0^\infty
\frac{dk}{2\pi}\overline{ e^{ik(X_{n_1}(\tau)-X_{n_2}(\tau))}}\\
\nonumber
&=&\int_0^\infty \frac{dk}{2\pi} e^{-k^2D\zeta|n_1-n_2|/2K}\\
&=&\sqrt{\frac{K}{2D\pi\zeta|n_1-n_2|}}
\end{eqnarray}
Inserting now \eq{gauss} into \eq{1timedelta} and performing the
integrations leads to
\begin{equation}
\label{I}
I=\frac{\Delta}{N^2\zeta^2}\sqrt{\frac{K}{2D\pi\zeta}}\int_0^t d\tau
\int_0^N dn_1 \int_0^N dn_2 |n_1-n_2|^{-1/2}=\frac{8\Delta}{3\zeta^2}
\sqrt{\frac{K}{2D\pi\zeta}}\frac{t}{\sqrt{N}}
\end{equation}
Using this expression in \eq{yp2} we see that
$\langle\overline{Y_n^2(t)} \rangle = D^N_{eff} t$, where now 
$D^N_{eff}$ is
\begin{equation}
\label{Deff2}
D^N_{eff}=\frac{D}{N}+\frac{4\Delta}{3\zeta^2}
\sqrt{\frac{K}{2D\pi\zeta}}\frac{1}{\sqrt{N}}
\end{equation}
Again the time dependent MdM flows are seen to produce a
``correction'' to the diffusion constant at long times; this
correction is of more importance the longer the chains.
For infinite chains $D^N_{eff}$ vanishes, a sign of the appearance for
$N\rightarrow\infty$ of the sublinear, $t^{3/4}$-behavior of \eq{bead2}. The
result can be understood in terms of 
the renewal of the process of \eq{bead2}: Thus for $t\gg
t_R\gg\Gamma$, assuming a loss of memory after
each $t_R$ units of time, $\langle\overline{Y^2_N(t)}\rangle$ can be
expressed as being around $t/t_R$ times $C\Delta/(\pi\zeta^2) 
(K/(4D^2\pi\zeta)^{1/4} t_R^{3/4}$. This leads to a correction of the diffusion
constant of  
\begin{equation}
\frac{1}{2t_R} \frac{C\Delta}{\pi\zeta^2}
\left(\frac{K}{4D^2\pi\zeta}\right)^{1/4}t_R^{3/4}=\frac{3C}{8\pi^{1/4}}
\frac{4\Delta}{3\zeta^2}
\sqrt{\frac{K}{2D\pi\zeta}}\frac{1}{\sqrt{N}}
\end{equation}
which is $3C/(8\pi^{1/4})$ times the correction in \eq{Deff2}.

A similar argument can be put forth when $t\gg\Gamma\gg t_R$. In this
case, $\langle\overline{Y^2_n(t)}\rangle$ can be viewed as arising
from renewals (every $\Gamma$ units of time) of the process given by
\eq{quench}. This leads to yet 
another effective diffusion constant $\tilde{D}^N_{eff}$, which
besides $D/N$ has an additional term proportional to 
\begin{equation}
\label{Deff3}
\frac{\Delta}{\zeta^2}\sqrt{\frac{N}{D\Gamma}}
\end{equation}
This reproduces (after replacing $D/N$ by $D$) the result of the
single particle case, \eq{Deff}.

We summarize the findings for the msd of a bead on the finite Rouse
chain. When $t\gg \max(t_R, \Gamma)$, there is normal but
anisotropic diffusion of the bead under 
observation, described by the diffusion constant
$D_{\perp}=D$ in the directions perpendicular to the flow,
and an effective diffusion constant $D_{||} > D$ parallel to the
flow. The value of $D_{||}$ depends on 
whether $t_R\ll\Gamma$ or $t_R\gg\Gamma$: For $t_R\ll\Gamma$, 
$D_{||}=\tilde{D}^N_{eff}$ (see \eq{Deff3}) and for $t_R\gg\Gamma$,
$D_{||}=D^N_{eff}$, \eq{Deff2}.
In the short time regime, $t\ll \min(t_R,\Gamma)$, we always have $Y_n^2(t)\sim
t^{7/4}$. However, the 
intermediate regime again depends on the magnitudes of $t_R$ and
$\Gamma$. If $t_R\gg\Gamma$ we observe
$\langle\overline{Y_n^2(t)}\rangle \sim 
t^{3/2}$ (\eq{quench}) in the regime $t_R\ll t\ll\Gamma$. But if on
the other hand $\Gamma\gg t_R$, then for times t obeying $\Gamma\ll
t\ll t_R$, the msd behaves as $\langle\overline{Y_n^2(t)}\rangle \sim
t^{3/4}$. Our model gives therefore rise to a very rich dynamical
behavior.

\section{Conclusion}
In this work we introduced a new variant of the Matheron-de Marsily
(MdM) model of layered random media, in which the velocity field is
also allowed to change with time. We solved exactly the equations
describing the dynamics of a single particle and of beads belonging to
infinite Rouse chains in time-dependent MdM flows. For short times and
slowly changing media, we recovered the 
results of the standard quenched MdM model. For large
times we found normal diffusion in the cases of a single particle and
of a finite Rouse chain, and we computed the corrections to the
diffusion constant. For the infinite Rouse chain we found a
surprisingly fast increase with time of the mean square displacement,
a result also valid for finite Rouse chains in intermediate time
regimes. 

\renewcommand{\abstractname}{Acknowledgements}
\begin{abstract}
The support of the DFG, of the GIF through grant I0423-061.14, and of
  the Fonds der Chemischen Industrie are gratefully acknowledged.
\end{abstract}

\end{document}